\documentstyle[twocolumn,epsf]{jpsj}
\def\runtitle{ 
Spin Excitation in Zigzag Nano-Graphite Ribbons
}
\def\runauthor{Hideo {\sc Yoshioka}}
\def\d{{\rm d}}

\def\e{{\rm e}}

\def\ggs{\buildrel\textstyle > \over {\hbox{\raise0.2ex\hbox{$\sim$}}}}
\def\lls{\buildrel\textstyle < \over {\hbox{\raise0.2ex\hbox{$\sim$}}}}
\def\gsim{\,\lower0.75ex\hbox{$\ggs$}\,}
\def\lsim{\,\lower0.75ex\hbox{$\lls$}\,}

\def\im{{\rm i}}
\def\ie{{\it i.e.}, }

\def\delx{\partial_x}
\def\deltau{\partial_\tau}

\def\bvec #1{\mbox{\boldmath $#1$}}
\def\ti #1{\tilde #1}
\def\e{{\rm e}}
\def\deltau{\partial_\tau}
\def\delx{\partial_x} 
\def\jo #1#2#3#4{#1 {\bf #2} (#3) #4}   %For J. Phys. Soc. Jpn.
\def\PRB{Phys.\ Rev.\ B}
\def\PRL{Phys.\ Rev.\ Lett.}

\def\APL{Appl.\ Phys.\ Lett.}

\def\JPSJ{J.\ Phys.\ Soc.\ Jpn.}

\title
{
Spin Excitation in Nano-Graphite Ribbons with Zigzag Edges
}

\author{
Hideo {\sc Yoshioka}\footnote{E-mail: yoshioka@phys.nara-wu.ac.jp} 
}

\inst{
Department of Physics, Nara Women's University, Nara 630-8506
}

\recdate{\hspace{3.5cm} }

\abst{
Spin excitation in a nano-graphite ribbon with zigzag edges is
investigated theoretically. 
Due to the strongly localized nature of the states near Fermi energy, 
the effective Hamiltonian for the low energy physics 
is given by Heisenberg Hamiltonian 
with the nearest neighbor exchange coupling. 
The action corresponding to the effective Hamiltonian 
is mapped to that of the $O(3)$ nonlinear sigma model. 
It is shown that the spin excitation has a gap when 
the number of the zigzag lines is even, whereas 
the excitation becomes gapless in case 
of the odd number of the zigzag lines.       
}

\kword{nano-graphite ribbon, spin excitation, nonlinear sigma model 
}

\begin{document}
\sloppy
\maketitle

%%%%%%%%%%%%%%%%%%%%%%%%%%%%%%%%%%%%%%%%%%%%%%%%%%%%%%%%%%%%%%%%%% 
Recently, the graphite-based one-dimensional materials with nano-meter sizes 
have been attracting much attention in both the fundamental sciences 
and the application sides. 
One of the most remarkable characteristics in these materials 
is that 
the electronic properties depend strongly on their geometrical structure.   
For example, 
the carbon nanotubes, which are made by rolling up a graphite sheet, 
can be either a metal or a semi-conductor  
depending on the chiral vector specifying the way of wrapping.\cite{Hamada,Saito-I,Saito-II} 
 
A nano-graphite ribbon (NGR) is the nano-meter size graphite fragment.
It has been clarified that 
the edge regions play important roles for  
the electronic structure in this material.
The calculation based on the tight-binding model revealed that 
a NGR with zigzag-shaped edges, 
which is called as a zigzag NGR in the following and shown in Fig.1, 
has strongly localized one-particle states around the edges.\cite{Fujita,Nakada,Wakabayashi} 
The same conclusion is obtained also by the first-principle calculation.\cite{Miyamoto}  
Since the localized states appear near the Fermi energy, 
it is expected that 
the mutual interaction between electrons affects the electronic states 
more strongly compared to the ordinary systems with 
extended states near the Fermi energy. 
Then  it may be necessary for clarifying the electronic properties to
utilize the different approach from the method for the ordinary system.    
   
The effects of the mutual interaction on the zigzag NGR 
have been investigated by applying the mean-field theory 
to  the Hubbard model with the on-site repulsive interaction and 
the hopping energy between the nearest neighbor atoms.\cite{Fujita,Wakabayashi-II}  
It has been shown that 
the infinitesimal on-site repulsion 
induces spontaneous magnetic moments; the conclusion is different
from that of two-dimensional graphite sheet, where 
the finite value is needed for appearance of the magnetic order.  
In addition, the large magnetic moments appear at the edges 
( 1A and NB sites in Fig.1 ).  
Based on this spin structure, the effective spin ladder model 
is proposed and the existence of the spin gap is
claimed.\cite{Wakabayashi-II}
However, since it is considered that, in one-dimensional systems, 
the effects of the mutual interaction 
are not sufficiently taken into
account in the mean-field treatment, 
the conclusion obtained by it has to be reexamined.
In the present work, we investigate the spin excitation of
the zigzag NGR with $N$ (number of the zigzag lines from one side to the
other, see Fig.1) 
based on the renormalization group analysis 
and by mapping the $O(3)$ nonlinear sigma model (NLSM). 
  Quite recently, 
Hikihara {\it et al.} investigated 
the ground state and the excitation around it 
of the zigzag NGR by both the bosonization theory and 
the density matrix renormalization group method.\cite{Hikihara} 
The bosonization theory insists that the ground state is a spin-singlet 
Mott insulator with finite charge and spin gap. 
The numerical calculation supports the conclusion in case of $N=2$.   
We compare our result with the above conclusions
obtained by the previous studies.

%----------- Fig.1 -----------
\begin{figure}
\vspace*{1em}
\centerline{\epsfxsize=8.0cm\epsfbox{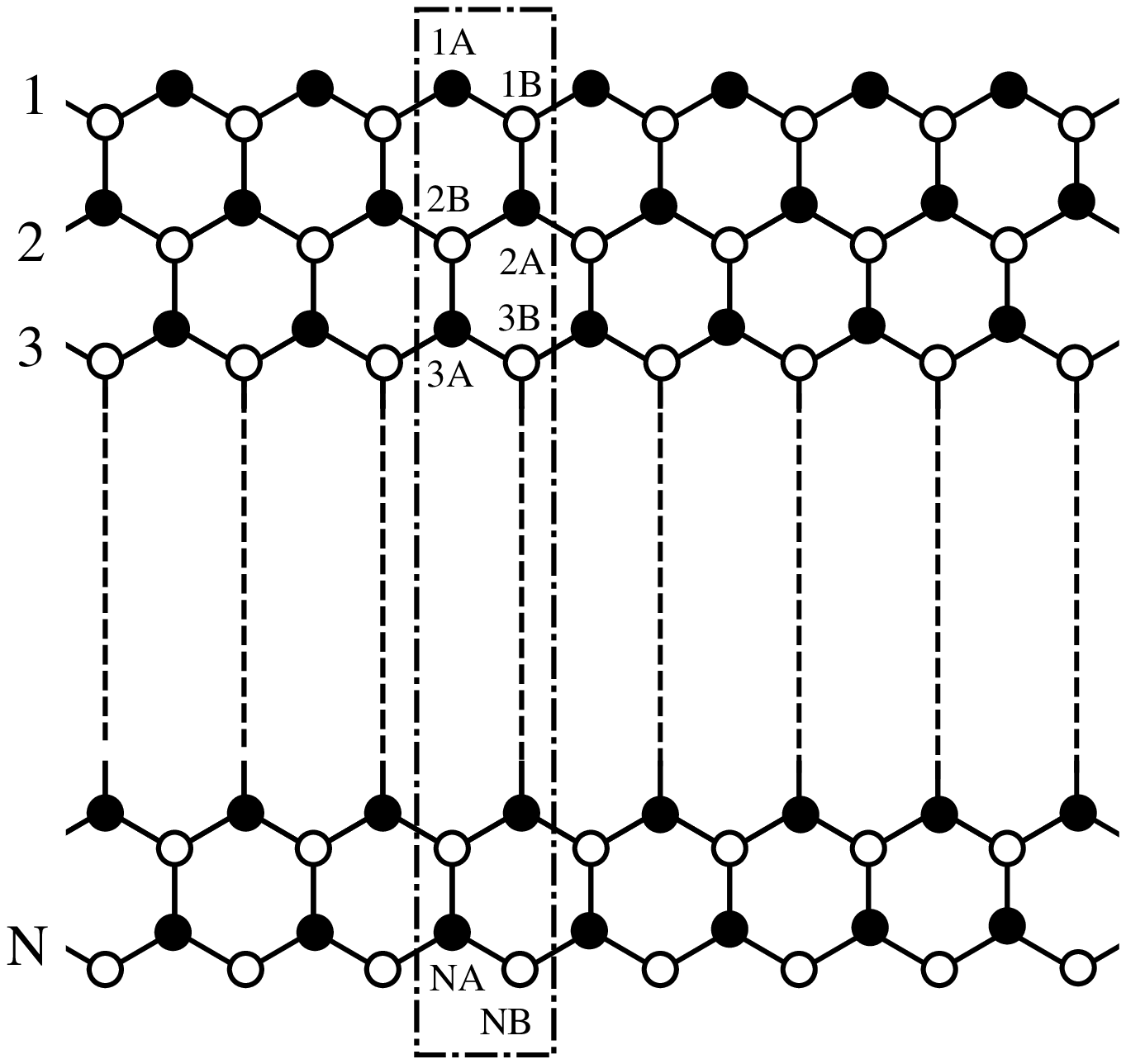}}
\caption{ The structure of the zigzag NGR with $N = {\rm odd}$. 
Here the close (open) circle shows the A (B) sublattice and 
the rectangle with the dashed-dotted line indicates the unit cell.
}
\label{fig:structure}
\end{figure}
%---------------------- 

\begin{figure}
\vspace*{1em}
\centerline{\epsfxsize=8.0cm\epsfbox{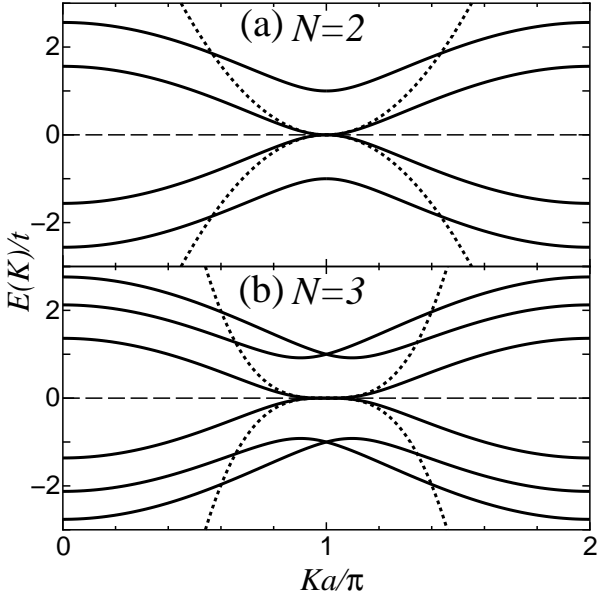}}
\caption{ The band structure of the zigzag NGR with $N=2$ (a) and $N=3$ (b). 
Here $t$ is the hopping energy between the nearest neighbor sites 
and $a$ is the lattice spacing.
The thin dashed curve express the Fermi energy and 
the asymptotic behaviors near Fermi energy, 
$E(K)/t \sim \pm (Ka - \pi)^N $ are written by the thick  dotted curves.   
}
\label{fig:band}
\end{figure}

As a model of the interacting electrons in the zigzag NGR, 
we consider the Hubbard Hamiltonian used in
refs. \citen{Fujita}, \citen{Wakabayashi-II}
and \citen{Hikihara}. 
At first we derive the effective Hamiltonian describing the low energy 
excitations.  
As was already discussed above,
the states near $K_F = \pi/a$ ($a$ : lattice spacing) 
have the strongly localized nature around the edges, 
and then 
the energy dispersion near $K_F$ are given as     
$E(K)\simeq \pm t (Ka - \pi)^N$ 
as was shown in Fig.\ref{fig:band}.\cite{Wakabayashi} 
%----------- Fig.2 -----------
%----------------------
When the effective low energy theory is constructed  
with taking account of the states near $K_F$, 
the interaction terms including the four fermion operators
have the scaling dimension, $N-1$, at the tree level as was already
discussed in ref.\citen{Hikihara}.  
Therefore, the interaction is always relevant for $N \geq 2$. 
In this case, one electron is localized at the one site 
and the system behaves as Mott insulator.  
Thus, the charge degree of freedom is frozen, \ie the charge excitation
becomes gapped, and 
the low energy physics is determined by the spin degree of freedom. 
In such a case, we should choose an approach from the atomic limit.   
Then, the effective Hamiltonian 
is given by the Heisenberg model with the nearest neighbor exchange
coupling $J$, 
which is written as follows,  
%------- (1-3) ---------
\begin{eqnarray}
{\cal H} &=& J \sum_{l=1}^{N_L} \left\{ h_{\rm odd}(l) + h_{\rm even}(l) \right\}, \\
h_{\rm odd}(l) &=& \sum_{i: \rm odd} 
\big\{ 
 \bvec{S}_{iA}(l)\bvec{S}_{iB}(l) 
+ \bvec{S}_{iB}(l-1)\bvec{S}_{iA}(l) \nonumber \\
& & + \bvec{S}_{iA}(l)\bvec{S}_{i-1B}(l)
\big\}, \\   
h_{\rm even}(l) &=& \sum_{i: \rm even} 
\big\{ 
 \bvec{S}_{iB}(l)\bvec{S}_{iA}(l) 
+ \bvec{S}_{iA}(l-1)\bvec{S}_{iB}(l) \nonumber \\
& & + \bvec{S}_{iA}(l)\bvec{S}_{i-1B}(l)
\big\}, 
\end{eqnarray}
%---------------------------------------
where $N_L = L/a$ with $L$ being the length of the NGR, 
$\bvec{S}_{0A/B}(l) = \bvec{S}_{N+1A/B}(l) = 0$ and   
$\bvec{S}^2_{iA/B}(l) = S(S+1)$  with $S=1/2$.
Though the magnitude of the spin, $S$, is a one-half, 
we do not restrict it in the following for convenience. 

The action $A = A_{\rm WZ} + A_0$ corresponding to the above Hamiltonian
is expressed by the coherent
state path integral as,\cite{Fradkin} 
%------------ (4-5)
\begin{eqnarray}
A_{\rm WZ} &=& \im S \sum_{l} \sum_{i=1}^N \left\{ w[\bvec{n}_{iA}(l)] + w[\bvec{n}_{iB}(l)]\right\}, \\
A_{\rm 0} &=& \frac{JS^2}{2} \int_0^\beta d \tau \sum_l
\Big[
\sum_{i : \rm{odd}} \big\{ \left(\bvec{n}_{iA}(l) + \bvec{n}_{iB}(l) \right)^2 \nonumber \\
&+& \left(\bvec{n}_{iB}(l-1) + \bvec{n}_{iA}(l) \right)^2 
+ \left(\bvec{n}_{iA}(l) + \bvec{n}_{i-1B}(l) \right)^2\big\} \nonumber \\ 
&+& \sum_{i : \rm{even}} \big\{ \left(\bvec{n}_{iB}(l) + \bvec{n}_{iA}(l) \right)^2 
+ \left(\bvec{n}_{iA}(l-1) + \bvec{n}_{iB}(l) \right)^2 \nonumber \\
&+& \left(\bvec{n}_{iA}(l) + \bvec{n}_{i-1B}(l) \right)^2\big\}
\Big],
\end{eqnarray}
%----------------------------------
where $\beta = 1/(k_B T)$ ($T$ : temperature) and   
the constraint $\bvec{n}_{iA/B}^2(l) = 1$ is imposed. 
The quantity, 
$w[\bvec{n}_{iA/B}(l)]$ is the solid angle which $\bvec{n}_{iA/B}(l)$ 
forms in the period $0 \leq \tau \leq \beta$. 
 
In the classical limit, 
the Hamiltonian (1)-(3) has the ground state with
$\bvec{S}_{iA}(l) = S \bvec{e}_z$ and $\bvec{S}_{iB}(l) = -S
\bvec{e}_z$ where $\bvec{e}_z$ is the unit vector along the $z$-direction. 
Here we discuss the fluctuation around the classical solution by 
expanding as $\bvec{S}_{iA}(l) = S \bvec{e}_z + \bvec{s}_{iA}(l)$
and $\bvec{S}_{iB}(l) = - S \bvec{e}_z + \bvec{s}_{iB}(l)$. 
The equations of motion of the fluctuation in the linearized
approximation are given as follows, 
%-------- (6-9) -------------------- 
\begin{eqnarray}
\dot{s}^+_{1A}(l) &=& -\im J S \left\{ 2 s^+_{1A}(l) + s^+_{1B}(l) + s^+_{1B}(l-1) \right\}, \\
\dot{s}^+_{iA}(l) &=& -\im J S \big\{ 3 s^+_{iA}(l) + 
s^+_{iB}(l) + s^+_{i-1B}(l) \nonumber \\
& & + s^+_{iB}(l+(-1)^i) \big\}, \hspace{2em}\mbox{for $i=2 \sim N$}, \\ 
\dot{s}^+_{iB}(l) &=& \im J S \big\{ 3 s^+_{iB}(l) + 
s^+_{iA}(l) + s^+_{i+1A}(l) \nonumber \\
& & + s^+_{iA}(l-(-1)^i) \big\}, \hspace{2em}\mbox{for $i=1 \sim N-1$}, \\
\dot{s}^+_{NB}(l) &=& \im J S \left\{ 2 s^+_{NB}(l) + s^+_{NA}(l) + s^+_{NA}(l-(-1)^N) \right\}, \nonumber \\
& & 
\end{eqnarray}
%------------------------------
where $s^+_{iA/B}(l) = s^x_{iA/B}(l) + \im s^y_{iA/B}(l)$. 
By introducing the Fourier transformation,
$s^+_{iA}(l) = (N_L)^{-1/2} \sum_k {\rm e}^{\im [k (l + (-1)^i/4)a -
\omega t]} s^+_{iA}(k)$ and
$s^+_{iB}(l) = (N_L)^{-1/2} \sum_k {\rm e}^{\im [k (l - (-1)^i/4)a -
\omega t]} s^+_{iB}(k)$, 
the eqs.(6)-(9) are rewritten as,
%-------- (10-13) ----------- 
\begin{eqnarray}
\ti{\omega} s^+_{1A}(k) &=& 2 s^+_{1A}(k) + 2 \cos (ka/2) s^+_{1B}(k), \\
\ti{\omega} s^+_{iA}(k) &=& 3 s^+_{iA}(k) + s^+_{i-1B}(k) + 2 \cos (ka/2) s^+_{iB}(k), \nonumber \\ & & \hspace{3em}\mbox{for $i=2 \sim N$}, \\  
\ti{\omega} s^+_{iB}(k) &=& - \left\{3 s^+_{iB}(k) + s^+_{i+1A}(k) + 2 \cos (ka/2) s^+_{iA}(k) \right\},
\nonumber  \\ & & \hspace{3em}\mbox{for $i=1 \sim N-1$},\\   
\ti{\omega} s^+_{NB}(k) &=& - \left\{2 s^+_{NB}(k) + 2 \cos (ka/2) s^+_{NA}(k) \right\}.
\end{eqnarray}
%------------------------------
The above equations have the two Goldstone modes whose frequencies are given
by $\ti \omega = \omega/JS = \pm 2 |\sin (ka/2)| \sim \pm |ka| \equiv
\omega_{\pm}(k)/JS$ (see Fig.3). 
Since the eigenvectors for the modes are obtained up to the $k$-linear as
%-------- (14,15)-------------- 
\begin{eqnarray}
 s^+_{iA}(k) &=& - ( 1 \pm \frac{N-i+1}{2}|ka|), \\
 s^+_{iB}(k) &=& 1 \pm \frac{N-i}{2}|ka|,  
\end{eqnarray}
%------------------------------
%----------- Fig.2 -----------
\begin{figure}
\vspace*{1em}
\centerline{\epsfxsize=8.0cm\epsfbox{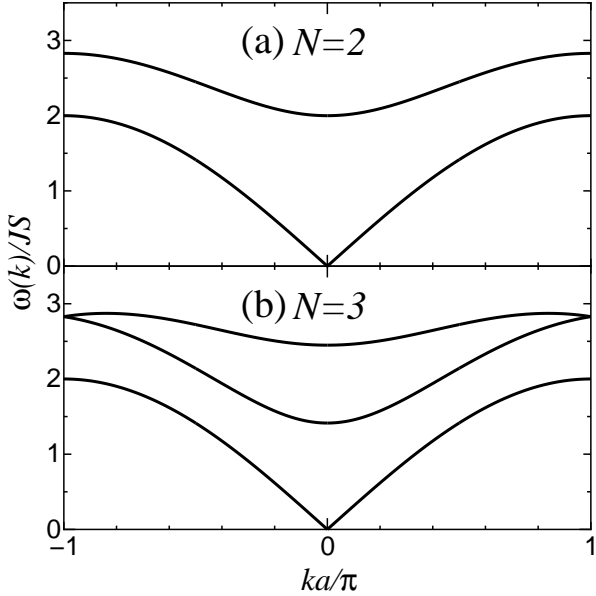}}
\caption{ The dispersion relation of the spin wave of the $N=2$ (a) 
and $N=3$ (b) zigzag NGR. 
Here the only the positive values are shown.  
}
\label{fig:SW}
\end{figure}
%----------------------
the fluctuation, $s^+_{iA/B}(l)$, can be written  as follows,  
%--------------- (16,17) --------------      
\begin{eqnarray}
s^+_{iA}(l) &=& \frac{1}{\sqrt{N_L}} \sum_k \e^{\im k (l + (-1)^i/4)a} 
 \Big\{(-1 - \frac{N-i+1}{2}|ka|) \nonumber \\ 
&\times& \alpha(k) + (-1 + \frac{N-i+1}{2}|ka|)\beta(k)  \Big\}, \\ 
s^+_{iB}(l) &=& \frac{1}{\sqrt{N_L}} \sum_k \e^{\im k (l - (-1)^i/4)a} 
 \Big\{(1 + \frac{N-i}{2}|ka|) \nonumber \\ 
&\times& \alpha(k) + (1 - \frac{N-i}{2}|ka|)\beta(k)  \Big\}, 
\end{eqnarray}   
%-------------------------------------------
where 
$\dot \alpha (k) = -\im \omega_{+} (k) \alpha (k)$ and 
$\dot \beta (k) = -\im \omega_{-} (k) \beta (k)$. 
When we introduce the spatially slowly varying functions made from the
Goldstone modes, 
%--------- (18,19) -------------- 
\begin{eqnarray}
 F(x_l) &=& \frac{1}{\sqrt{N_L}} \sum_k \e^{\im k (l - 1/4)a} \nonumber \\
&\times& \left\{-(1 + |ka|/4) \alpha(k) - (1 - |ka|/4) \beta(k)\right\}, \\
 G(x_l) &=& \frac{1}{\sqrt{N_L}} \sum_k \e^{\im k (l - 1/4)a} 
 \left\{-|ka|/2 ( \alpha(k) - \beta(k) )\right\} , \nonumber \\
& & 
\end{eqnarray}
%-------------------------------
with $x_l = l a$, 
eqs.(16) and (17) are written as follows, 
for $i=$odd, 
%------- (20,21) -----------
\begin{eqnarray}
s^+_{iA}(l) &=& F(x_l) + (N-i+1/2) G(x_l), \\
s^+_{iB}(l) &=& -F(x_l+a/2) - (N-i-1/2) G(x_l+a/2), \nonumber \\
& &  
\end{eqnarray}
%----------------------
and for $i=$even, 
%------- (22,23) -----------
\begin{eqnarray}
s^+_{iA}(l) &=& F(x_l+a/2) + (N-i+1/2) G(x_l+a/2), \nonumber \\
& & \\
s^+_{iB}(l) &=& -F(x_l) - (N-i-1/2) G(x_l) .
\end{eqnarray}
%------------------------ 

Next,  we try to map the action given by eqs.(4) and (5) to that of the
$O(3)$ NLSM. 
Eqs. (20)-(23) suggest the following ansatz, for $i$=odd,
%--------- (24,25) ----------- 
\begin{eqnarray}
\bvec{n}_{iA}(l) &=& \bvec{\Omega}(x_l) + (N-i+1/2) \frac{a}{S} \bvec{l}(x_l), \\
\bvec{n}_{iB}(l) &=& - \bvec{\Omega}(x_l+a/2) - (N-i-1/2) \frac{a}{S} \bvec{l}(x_l+a/2), 
\nonumber \\ 
& & 
\end{eqnarray}
%-------------------
and for $i=$even,
%--------- (26,27) -----------  
\begin{eqnarray}
\bvec{n}_{iA}(l) &=& \bvec{\Omega}(x_l+a/2) + (N-i+1/2) \frac{a}{S} \bvec{l}(x_l+a/2), \nonumber 
 \\ & & \\
\bvec{n}_{iB}(l) &=& - \bvec{\Omega}(x_l) - (N-i-1/2) \frac{a}{S} \bvec{l}(x_l).
\end{eqnarray}
%-------------------------
The constraint $|\bvec{n}_{iA/B}| = 1$ leads to 
$|\bvec{\Omega}| = 1$ and $\bvec{\Omega} \bvec{l} = 0$ 
to the first order of $a/S$.  
By substituting eqs.(24)-(27) into eq.(4), 
we obtain in the continuum limit,
%---------- (28) -----------------
\begin{eqnarray}
 A_{\rm WZ} &=& \im \theta Q + \im N \int \d \tau \d x (\deltau \bvec{\Omega} \times \bvec{\Omega}) \bvec{l}, 
\end{eqnarray}
%--------------------------------
where $\theta = 2 \pi S$ for $N =$odd and $\theta = 0$ for $N=$even. 
The quantity $Q$ is the winding number of the Euclidean space
configuration $\left\{ \bvec{\Omega}(x,\tau)\right\}$ and written as,
%----------- (29) -------------------- 
\begin{eqnarray}
 Q &=& \frac{1}{4 \pi} \int \d \tau \d x (\deltau \bvec{\Omega} \times \delx \bvec{\Omega}) \bvec{\Omega}.
\end{eqnarray}
%----------------------------
On the other hand, the quantity $A_0$ is expressed as 
%---------------- (30) --------------- 
\begin{eqnarray}
A_0 &=& \frac{J S^2 N a}{2} \int \d \tau \d x \left\{ \frac{1}{2} (\delx \bvec{\Omega})^2 + \frac{2}{S^2} \bvec{l}^2\right\}.
\end{eqnarray}
%---------------------------------------
By integrating the field $\bvec{l}$ in eqs.(28) and (30), 
we obtain the following effective
action of the NLSM as
%----------------- (31) ------------------ 
\begin{eqnarray}
 A_{\rm eff} &=& \im \theta Q + \frac{1}{2g} \int \d^2 x (\partial_\mu \bvec{\Omega})^2 ,
\end{eqnarray}
%----------------------------------------
where $x_1 = x$, $x_2 = v \tau$ with $v = JSa$ and 
$g = 2/(NS)$. 
The above effective action is the same as that for spin ladder
systems\cite{Dagotto,Senechal,Aringa} and 
the excitation is gapless (gapped) for $N=$ odd (even) in case of $S = 1/2$. 
The conclusion is different from the result 
based on the mean-field theory and that by the bosonization theory 
where there is the spin gap irrespectively of the number, $N$.
The discrepancy is considered to be due to the following facts. 
The effects of the interaction are difficult to be taken into account
sufficiently  in the mean-field theory starting from the band picture 
and the consistent mapping from the Fermionic theory to the bosonic one 
is not done in the present case without the linear dispersion across 
the Fermi energy.  
In ref. \citen{Hikihara}, the spin excitation has been calculated for
$N=3$ numerically.
It seems to us that   
the result indicates that the excitation for $N=3$ is gapless, 
which is identical with our conclusion,  
though the authors do not insist it because of lack of the enough system
size for extrapolation.  
   
We note that the velocity of the excitation $v$ 
is identical with that obtained by the spin wave analysis. 
In addition, in case of $N=1$, 
the present results are identical with that with one-dimensional spin
chains\cite{Affleck} with $a \to 2 a$ because the lattice spacing 
becomes a half.

The coupling constant, $g$, for $N=2$ obtained here is smaller than that 
for the two-leg spin ladders with the isotropic coupling.\cite{Senechal,Aringa} 
Then the spin gap of the present system is expected to be smaller than 
that of the spin ladders. 
This fact is considered to be due to the lack of the exchange coupling 
between {\it eg}., 1A and 2B. 
In deriving the effective action, eq.(31), the modes with high frequencies 
have been neglected.    
We have derived the effective action with taking account of 
both the Goldstone modes and the high frequency modes in case of $N=2$.  
The conclusion is the same as eq.(31) with $N=2$. 
Thus, it is expected that the high frequency modes do not play an 
important role for the analysis presented here.       

In summary, we investigated the spin excitation of the zigzag NGR with the
width $N$. 
We concluded that the spin excitation is gapless for $N=$ odd, 
whereas the excitation has a gap  in case of $N=$ even.  

%------------Acknowledgement--------
\section*{Acknowledgments}
The author would like to thank T. Hikihara for valuable discussion and comments.
This work was supported by Grant-in-Aid  for Scientific Research (A)
(No. 13304026) and (C) (No. 14540302)
from the Ministry of Education, Culture, Sports, Science and Technology,  
Japan.
%===============================================================

\end{document}